\documentstyle[prb,aps,twocolumn]{revtex}
\begin{document}

\draft

\title{Study of theoretical models for the liquid-vapor and metal-nonmetal
transitions of alkali fluids }

\author{
E. Chac\'on$^1$ ,
J. P. Hernandez$^2$, and
P. Tarazona$^3$ }

\address{
$^1$Instituto de Ciencia de Materiales, Consejo Superior de
Investigaciones Cient\'ificas and \\  Departamento de
F\'isica Fundamental, Universidad Nacional de Educaci\'on a
Distancia, Apartado 60141, E-28028 Madrid, Spain}

\address{
$^2$Department of Physics and Astronomy, University of North Carolina,
Chapel Hill NC 27599-3255 USA }

\address{
$^3$Departamento de F\'isica de la Materia Condensada (C-XII),
Universidad Aut\'onoma de Madrid, E-28049 Madrid, Spain }

\date{\today}
\maketitle

\begin{abstract}
Theoretical models for the liquid-vapor and metal-nonmetal transitions of
alkali fluids are investigated. Mean-field models are considered first
but shown to be inadequate, apparently due to their inability to allow a
microscopically consistent treatment of coexisting localization and
delocalization of the valence electrons in the materials. An alternate
approach is then studied in which each statistical configuration of the
material is treated as inhomogeneous, with the energy of each ion being
determined by its local environment. Nonadditive interactions, due to
valence electron delocalization, are a crucial feature of the model. This
alternate approach is implemented within a lattice-gas approximation which
takes into account the observed mode of expansion in the materials of
interest (a change in the average coordination rather than a change in the
average nearest-neighbor distance) and which is able to treat the equilibrium
density fluctuations. We have carried out grand canonical Monte Carlo
simulations, for this model, which allow a unified, self-consistent, study
of the structural, thermodynamic, and electronic properties of alkali
fluids. Applications to Cs, Rb, K, and Na yield results in good agreement
with experimental observations.

\end{abstract}

\pacs{ PACS numbers: 61.25.Mv, 64.70.Fx, 71.30.+h }

\section*{Introduction}

        Until recently little has been known regarding the interrelation of
the structural, thermodynamic, and electronic properties of metal-atom
fluids. However, such knowledge has been being developed in a substantial
body of experimental data which has been crying out for theoretical
interpretation and guidance. The electronic, structural, and thermodynamic
properties of such fluids have been shown to be intimately related and the
interdependence of the metal-nonmetal and liquid-vapor transitions has
posed a challenge to theoretical understanding. Detailed experimental
studies are available for those materials with the lowest liquid-vapor
critical temperatures: Hg (1751 K), Cs (1924 K), and Rb (2017 K) \cite
{Gener1}, with data on K (2178 K) becoming available most recently \cite
{Hohl}. The data have become precise and reliable in the last decade and
span thermodynamic and electrical measurements under the same conditions.
Such data show that the liquid-vapor coexistence curve of metal-atom fluids
are different from those of Lennard-Jones-like ones \cite {Hohl,Jungst},
presumably due to many-body effects associated with valence electron
delocalization. For example, the law of corresponding states is not obeyed
when metal-atom fluids are compared with pair-interacting ones. Also, the
liquid and vapor branches of the coexistence curves are strongly asymmetrical
and the rectilinear diameter law breaks down over a substantial temperature
range, not only very close to the critical points. These materials also
undergo a metal-nonmetal (M-nM) transition. This body of data, however,
still seeks microscopic theoretical foundations \cite {Stratt}. This paper
presents a study of theoretical approaches seeking to comprehensively
understand the alkali fluids. After demonstrating that a series of
approaches, which would seem appropriate, fail to explain the general
features observed, a simple model is described which does appear to contain
the basic ideas required to reproduce, in a unified manner, the peculiar
characteristics observed in the alkali fluids. This paper presents a
detailed treatment of our results, improving and detailing the
information in our recent letter \cite {Tarazona}.

        The goal of the present work -- a unified understanding of the
structural, thermodynamic, and electronic properties of metal-atom fluids --
poses a considerable scientific challenge. Its various aspects are coupled
since it is the electrons which determine interatomic interactions and thus
the material structure and thermodynamic data. The ionic structure, in
turn, determines the electronic properties. In the study of metal-atom
fluids it is difficult to impose a structure, since it is so intimately
related to electronic effects and there is no long-range symmetry to
simplify the problem. Also, because in such materials the interactions are
not pairwise ones for large and intermediate densities, due to electron
delocalization over some regions, the problem is more complicated than that
for simple, Lennard-Jones-like, fluids in which the interactions are not state
dependent. Similarly, as these materials also undergo a M-nM transition,
the traditional techniques used to study free-electron-like fluids are not
applicable over many of the conditions of interest. The microscopic theory
required for these materials should seek to explain the essential
interdependence of thermodynamic, structural, and electronic properties.

Previous theoretical efforts to comprehensively explain the available
experimental data on metal-atom fluids have been sparse. The points of view
taken were usually based on the limiting cases of either a metallic, dense,
liquid or solid, or alternatively of a nonmetallic, dilute, vapor. Attempts
were then made to describe the fluid, or some of its properties, in a
limited density and temperature range \cite {Hafner,Minchin,Hernan1}.
General arguments based on electron correlation effects (Hubbard model)
and/or disorder induced localization (Anderson model) \cite {Lee} are
useful to study the M-nM transition in systems with frozen ionic structure
but probably not in metal-atom fluids; at least, we are unaware of
calculations attempting to link the structural, thermodynamic, and
electronic properties using such methods.

A first step towards a theoretical treatment of metal-atom fluids, which
is intended to apply to both low and high densities at temperatures
above approximately 2000 K, extended concepts and techniques of plasma
physics, in a mean-field approach, and introduced the required neutral
atoms and small clusters \cite {Gener2}. These authors have recently
claimed partial quantitative success in linking their calculations and a
spectrum of experimental data. We also proposed such an approach \cite
{Chacon,Hernan2}, not limited to high temperatures and including a
discussion of the M-nM transition. We showed that a toy model gives a
liquid-vapor coexistence and a critical point with some correct features,
but we were far from reproducing the peculiar coexistence curve of alkali
fluids. In preliminary work \cite {Rostoc}, we followed this approach
including a quantitatively good description of the charged particle system:
a standard description of liquid metals near their melting point \cite
{Shimoji}. Extension of this treatment to high temperature and low density
gave a liquid-vapor coexistence with a very high critical
temperature (around four times the experimental value), very low critical
densities and pressures (by about an order of magnitude, compared to
experiment), and a very different shape for the coexistence curve than that
observed. To deal with the M-nM transition, we then extended the model,
using a statistical treatment, to allow for chemical coexistence of neutral
atoms with the ions. Phenomenological ion-atom interactions have been
used, in work to be reported below, instead of the neglect of atom
interactions assumed in the previous toy model. However, reasonable
values of the parameters have not improved the previous results, in
contrast to claims by others \cite {Gener2}. Details of this approach will
be given in this paper. We have concluded that a mean-field theory is not
capable of reproducing the structure or phase diagram of the alkali
fluids. The physical reason for this failure is discussed below.

An unanswered question in a mean-field approach, with an atomic and a metallic
component, is: Why do some fraction of the valence electrons choose to be
bound in atoms while others are delocalized, at fixed temperature and
chemical potential? The answer must lie in a hitherto ignored underlying
structure. An important clue is that clustering effects are strongly
enhanced for metal-atom systems, compared to nonmetallic ones, due to their
high cohesion, which arises from the valence electron delocalization over the
cluster. In contrast, to retain its valence electron an atom should have no
near neighbors to which that electron can be favorably delocalized.
Further, experimental neutron scattering data \cite {Winter} have shown
that the materials expand by changing their average coordination,
rather than their average nearest-neighbor (nn) distance. This fact is
especially important in systems in which valence electrons delocalize since
such delocalization leads to contributions to cluster energies which go
beyond merely additive effects in the local coordination; metal-atom
clusters are strongly bound and their cohesion is a non-linear function of
their density. Thus, in contrast to a mean-field characterization,
structural and electronic effects are intimately related when equilibrium
density fluctuations can be appreciable; this is the case in the expanded
metal-atom liquid and vapor cases.

Based on the above, we have explored a model which takes into account, from
the beginning, that expansion in the materials of interest takes place
through a change in the average coordination, rather than by changing the
nn distance, and which allows treatment of equilibrium density
fluctuations. We have begun with the simplest model which seeks to give a
recipe for the energy of an ion in a specific local environment, including
effects due to possible valence electron delocalization. The recipe is to
treat each local environment as a macroscopic one with an average density
equal to that due to the ion in question and its local coordination. The
possibility of valence electron localization, to form an atom, is taken
into account by choosing the lower energy arising from treating the system
as a metal of low density or an atom, dimer, etc. By treating each
statistical configuration of the material as inhomogeneous, such a recipe
provides the basis with which a self-consistent treatment of the material
structure may be sought. To implement the self-consistent structural
treatment, we have begun with a lattice-gas approximation and obtained its
equilibrium properties using grand canonical Monte Carlo (MC) simulations.
This approach yielded results showing the observed peculiarities of the
alkali fluids \cite {Tarazona}. That work has been improved and full
details will be given here. Results for Cs, Rb, K, and Na have been
obtained and will be presented.

To orient the reader, we summarize the contents of this paper. In section
\ref{sec:mean} we discuss the mean-field treatment, for its own sake and
with a view to an application discussed later. We begin discussing a
hypothetical metal at an assumed arbitrary density. Thermal effects due to
the delocalized electrons are examined. Pseudopotential parametrizations
are discussed and compared. Hard-sphere and one-component-plasma
reference systems for the ions are compared and contrasted. Then, atoms are
introduced in thermal equilibrium with the metallic component. Results,
conclusions, and criticism of this approach follow. The crux of the
model which proves successful follows. Section \ref{sec:model}
discusses our approach and its implementation within the lattice-gas
approximation. The grand canonical Monte Carlo treatment of the lattice gas
follows; results of its application to the spectrum of alkali fluids are
then given. Phase coexistence, structural features, and conductivity results
are displayed and discussed. The paper concludes with a summarizing discussion
and suggestions for further work.

\section{Mean-field treatment}
\label{sec:mean}
\subsection{Metal}

At first, we consider a hypothetical system, entirely composed of
ions and a neutralizing sea of valence electrons; this system exhibits a
liquid-vapor transition. Atoms, with their localized valence electrons, in
thermal equilibrium with the ions and delocalized electrons will be added
as a next step.  A mean-field treatment for a system of positive ions, at
a mean density $ \rho $, and delocalized, neutralizing electrons is the
normal one used for liquid alkali metals \cite {Shimoji}. It is based on
the Gibbs-Bogoliubov inequality:
\begin{eqnarray*}
f \leq f_o + < H - H_o >_o / V = f_o + u ( \rho ) \ \ ,
\end{eqnarray*}
where $ f $ is the free energy per unit volume $ V $, $ H $ is
the system Hamiltonian, and the subscript indicates an ionic
reference system. For the present problem we use an ionic reference system
and a jellium treatment of the delocalized valence electrons, a
pseudopotential $ v _{ps} $ is then associated with the ions and the
screened ion-electron and ion-ion interactions (minus reference system
effects) are treated by perturbation theory. Hence, we can write:
\begin{eqnarray*}
u ( \rho )  = u_{si} ( \rho _e ) + \\
2 \pi \rho ^{2} \int _{0} ^\infty
dr r ^2 g_0 ( r ; \rho ) \phi (r ; \rho_e ),
\end{eqnarray*}
where $ \rho _e = Z \rho $ is the electronic density, $ u_{si} $ is the
sum of all structure-independent terms including the kinetic, exchange and
correlation electronic energies of the jellium reference system and the
first order pseudopotential perturbation term; $ \phi (r; \rho _{e} ) $ is the
total interatomic potential which is given by:
\begin{eqnarray*}
\phi ( | {\bf R} - {\bf R'} |; \rho_e )  =  \frac { Z e ^{2} }{
| {\bf R} - {\bf R'} | }  +  \\    \int d {\bf r} d {\bf r'}
 v _{ps} ( | {\bf r} - {\bf R } | ) \chi (
| {\bf r} - {\bf r'} |; \rho _e )
v _{ps} ( | {\bf r'} - {\bf R'} | ),
\end{eqnarray*}
where $\chi$ is the electronic linear response function. For the ions, a
hard-sphere (HS), or alternatively a one-component-plasma (OCP), reference
gives entropy contributions $f_o $ and a pair distribution function $ g_{o}
( r ; \rho ) $. If using the HS reference for the ions, one makes use of
the Gibbs-Bogoliubov inequality by choosing the ion HS radius to minimize
the system free energy at the density and temperature which correspond to
the melting point for the liquid metal. This radius can also be used to
minimize the free energy at higher temperatures, though the change in the
obtained radius is negligible except at high densities. In the OCP choice,
the ionic system is described via the parameter $ \Gamma = (4 \pi \rho /
3)^{1/3} e^{2} /k_B T $, which is not treated as a variational parameter.
The free energy estimates then allow discussion of the liquid-vapor phase
diagram of the hypothetical material. Model results can be judged in a
partial manner, here and later, by their approximate predictions for the
vapor-liquid critical parameters; when such parameters are reasonable, a
more detailed comparison with experiment can be sought.

In our calculations, thermal effects on the delocalized electrons are not
taken into account. However, these effects had been probed in calculations
of the phase diagram of a delocalized electron system, neutralized by
jellium, in the temperature-dependent Hartree-Fock approximation \cite {HF};
an ideal-gas reference for the ions was also incorporated. Such a treatment
yields a critical point at 2550 K with a density of $ 1.3 \times 10^{-4}$
(a.u.). Replacing the temperature-dependent electron gas with a
zero-temperature one, and maintaining exchange as the only net
interaction, resulted in an increase of about 100 K in the critical
temperature, a near-negligible correction. Further, if the correlation
energy due to the zero-temperature electron gas is incorporated into such a
calculation, the critical point temperature rises substantially. An ion
ideal gas added to a zero-temperature electron gas, which includes
correlations, gives critical parameters of 4184 K and $ 1.8 \times
10^{-4}$. Use of Pad\'{e} approximants \cite {Gener2}, for the finite
temperature electron gas, only changes the previous coexistence
curve slightly, with the critical conditions then being given by 4500 K and
$ 1.6 \times 10^{-4}$. Such, ten per cent, corrections suggest that thermal
effects due to the electron gas are not important to the calculation of
near-critical conditions in these systems.

Pseudopotential effects are then required. In the study of alkali liquid
metals near their melting point, it has been common to simplify the ionic
pseudopotential, which characterizes the material, by choosing one which is
local and energy independent. In our previous calculations \cite {Tarazona}
we used the empty-core  Ashcroft pseudopotential. Here we investigate the
differences which arise from that choice and a Shaw pseudopotential
(constant in the core and continuous at the cutoff radius). Both
pseudopotentials only depend on one parameter, which is chosen to fit some
experimental result for the liquid at the melting point, and then checked
against a set of other experimental features. The set of experimental data
of interest at the melting point consists of: the packing fraction (since
the hard-sphere diameters are fixed to minimize the energy at the
experimental density), the ionization potential of the atom (as a
pseudopotential should reproduce the energy spectrum of states outside the
core), the cohesive energy, the electrical conductivity (using the Ziman
formula), the excess entropy over that of the ideal gas, and  the
pressure (one atmosphere). A common treatment of liquid metals \cite
{Shimoji} uses the Ashcroft pseudopotential fitted to the conductivity;
many of the parameter choices reflect a fit to the conductivity in the
solid. However, for some liquid alkali metals, such parameter choices yield
a nonphysical packing fraction and are then reparametrized to fit a packing
fraction of 0.45. This was the parametrization we used in previous work
\cite {Tarazona}. However, in attempting to deal with the alkali family,
rather than only one or two alkalis, we decided that fitting the presure at
the melting point should be preferable. Also, we noticed that the previous
parametrizations tended to substantially overestimate the cohesive energy
of the liquids (relative to separated atoms) at the melting point. After
examining the results of such parametrizations, and requiring that
adjustments of the parameter to fit one property not result in spoiling the
agreement with others, we have opted to use the alternate Shaw approach,
fitted to the experimental pressure at the melting point. The ionization
potentials, the cohesive energies, and the pressure are improved and we
believe that such improvements are important for our purposes. Table I
compares the results of the Ashcroft pseudopotential (fitted to the packing
fraction or conductivity) and the Shaw alternative (fitted to the pressure)
when applied to the liquid alkali metals at the melting point. We shall later
compare results we have obtained using both approaches in the present
application.

We have also explored the relative effects of the HS and OCP reference
systems for the ions. There are two features which have an influence
on the results obtainable: the excess entropy beyond the ideal gas,
and the pair distribution function which enters in calculating the screened
pseudopotential contributions. We have found that the excess entropy
term is substantially different in the two choices of an ionic reference
system, while the pair distribution and associated interaction energies are
not a strong cause of differences between them. One must be careful in the
comparison to note that, if the screened pseudopotential terms are
excluded, the HS system only has the excess entropy but the OCP has such a
term and the classical interaction of the ions in the electron sea. To
quantify the effects of the ionic reference system, we have performed a
series of calculations. We first sought the temperature and density of the
critical point of an unspecified metallic system, excluding the screened
pseudopotential terms, with the electron system treated at zero temperature.
The HS reference yielded: 4200 K  and a density  of $ 1.8 \times 10^{-4} $;
it should be noted that these results are almost identical to those using
an ideal-gas reference, thus the excess entropy of HS over the ideal gas has a
very small effect on the critical parameters, the density being so small. The
OCP, however, yielded: 10600 K and $ 3.2 \times 10^{-3} $; the strong
effect of the ionic interactions in the OCP leads to the higher critical
temperature. Then, the effects of the screened pseudopotential were
introduced. The results are then as follows, if one uses parametrizations
with an empty-core and a cut-off radius of 1.70 a.u. (appropriate for Na).
For HS, the calculated critical parameters change to 9000 K and $ 6 \times
10^{-5} $. For the OCP, the results become 4600 K and $ 1.5 \times 10^{-4}
$. The interactions raise the HS critical temperature, while the partial
cancellation of the classical ion interactions lowers that appropriate to
the OCP. Since, recalling the Gibbs-Bogoliubov inequality, the free energy
estimates are an upper bound to the correct ones, obviously the better
reference system is that which yields a lower value for the free energy.
Hence, it is of interest to compare the free energies themselves,
instead of focusing on a comparison of the above results to experimental
data.  Fig. 1 shows the free energy difference, with the sodium
parametrization as an example of the result typical of all alkalis, for HS
minus OCP as a function of density for three temperatures. It is clear that
for the density range of interest here, $ \rho > 10^{-5} $, the HS
reference system is to be preferred, having a substantially lower free
energy of one to three eV per particle. A criticism of previously noted work
\cite {Gener2} is that it uses the OCP reference system.

In table II we present the critical parameters, for the hypothetical alkali
metals (labeled as MF-hs), obtained with the Shaw parametrization, the HS
reference system, and the electronic system treated at zero temperature. In
fig. 2 we show the coexistence curve for rubidium, obtained with
the Shaw and the Ashcroft parametrizations; the difference here is relatively
small. Also, it is clear that this type of result is in poor
agreement, quantitatively and qualitatively, with experimental observations
\cite {Gener1,Hohl,Jungst}. These results are typical of all alkalis.
However, there are no atoms in the system, an important feature according
to previous work \cite {Gener2,Chacon} and crucial to incorporate the M-nM
transition observed.

Since, physically, at moderately low densities the system must have the
valence electrons localized on the ions but delocalized at high densities,
the valence electrons must coexist in localized and delocalized states at
fixed total chemical potential, pressure, and temperature, for phase
coexistence. Thus, we proceed to consider a thermodynamic equilibrium
mixture. We will restrict the mixture to a metallic component, as above,
and atoms, without further complicating the treatment by considering
diatomic molecules or other types of aggregates in this mean-field approach.

\subsection{Equilibrium with atoms}

The general approach followed here is along the lines previously noted
\cite {Chacon}. The procedure uses an approximation to the system free
energy in which ideal-gas terms for ions and atoms are first explicitly
separated, the remainder of the free energy is denoted as $f _{e}$:
\begin{eqnarray*}
f \equiv   k _{B} T [ \rho _{a} ( ln ( \rho _{a} \Lambda ^{3} /2 ) - 1 ) +
\rho _{i} ( ln ( \rho _{i} \Lambda ^{3} ) - 1 ) ] + \\
f _{e} ( \rho _{i}, \rho _{a}, T ).
\end{eqnarray*}
The atom and ion densities are denoted by $\rho$, with appropriate
subscripts, and $\Lambda$ is their thermal de Broglie wavelength. Then, the
grand potential per unit volume is given by:
\begin{eqnarray*}
\Omega / V  = f - \mu _{a} \rho _{a} -
\mu _{i} \rho _{i} = \\
f - \frac {E}{2} ( \rho _{a} - \rho _{i} )  -
\frac {  \mu _{a} \rho _{a} -
 \mu _{i} \rho _{i} }{2}  ( \rho _{a} + \rho _{i} ),
\end{eqnarray*}
where $\mu$ are the chemical potentials and E is the vacuum ionization
potential of the atom in question. $\Omega$ is first extremized with respect
to the difference between atom and ion densities, with the sum of the
densities being kept constant. The result of this procedure is an equilibrium
relation between atom and ion densities:
\begin{eqnarray*}
\rho _{a} = 2  \rho _{i} exp[(\mu _{e} + E)/k _{B}T] ,
\end{eqnarray*}
where $ \mu _{e} = \partial f _{e}/\partial \rho _{i} - \partial f
_{e}/\partial \rho _{a}$. Using this equilibrium relation then leaves the
sum of atom and ion densities and the temperature as the only variables
which can yield a liquid-vapor transition of the composite system. A jump
in the relative ion to atom density, at fixed sum of densities,
characterizes a resulting, first-order, M-nM transition, but this
transition, and its critical point, can only take place hidden under the
liquid-vapor phase coexistence of the composite system.

At first we considered a free energy with contributions from the electron
gas, a neutralizing ionic system, and an ideal gas of atoms, with the atoms
experiencing no interactions with the charged-particle system; thermal
equilibrium with the metallic component was demanded. The work previously
reported \cite {Chacon} showed that such an introduction of atoms would
merely shift the critical point to higher density and pressure, from the
conditions obtained without the atoms, without affecting the critical
temperature. Although that work included only the electronic kinetic energy
and exchange in $f_{e}$, the addition of correlation energies and the
ionic terms would leave the above conclusion unchanged: the introduction of
an ideal gas of atoms in thermal equilibrium with the charged particle
system yields a liquid-vapor critical temperature which is that due to the
system of charges and is independent of the ionization potential of the atoms.
It has already been noted that such a temperature is extremely high,
compared to the experimental results.

We proceeded, in a modification appropriate to Cs, by merely adding a HS
reference for atoms with a diameter (10 a.u.) different than that of the ions
(previously noted to be 8.80 a.u.); we used the method of
Mansoori et al \cite {Mansoori}. The dashed line of fig. 3 shows the
coexistence curve obtained. A very high critical temperature resulted, and
moreover the shape of the coexistence curve became clearly unphysical:
with a density for the vapor branch nearly independent of the temperature
and pinned by the mean-field, first-order, M-nM transition line. We
conclude that only atom-atom and atom-charge interactions can change the
critical temperature appreciably from that due to the charged particle
system. The effect of such terms is to influence the relative populations
of the atomic and metallic components in the system, as thermodynamic
equilibrium between them is demanded.

As the interactions due to atoms tended to affect the coupling between the
M-nM and liquid-vapor transitions, our main purpose in this investigation
was to observe the influence of the atom interactions on the liquid-vapor
critical point of the system. To carry out this investigation in full
detail is quite complicated; for example, to include the polarization of
atoms by the charged particles requires calculation of a micro-electric
field distribution. The polarization energy of a neutral atom arises from
the square of the sum of the (vector) electric fields to which it is
exposed; it cannot be written as the sum of interaction energies with the
individual charged particles (in contrast to previous assumption \cite
{Gener2}). Given this complication and also the fact that a pseudopotential
treatment of atom-electron interactions has doubtful validity (as it treats
the localized valence electron as frozen), we decided to begin by
investigating a phenomenological treatment for the atom-charges
interactions: a sum of a constant coefficient times the square of the atom
density plus a second constant times the product of the atom density and a
power of the charged particle density; the second term would be of dominant
importance. Various exponents in the second term were considered.

Treating the atoms as hard spheres, with a diameter different than that of
the ions, and further including (in $f_{e}$) their virial interaction with
the charges, $ a \rho_{a} \rho_{i}$, gives a first example of the type of
effects obtainable. We used the empirical prefactor $a$ as a free
parameter and analyzed the changes it induced in the critical temperature.
This temperature, as a function of $a$, has a minimum which is
less than five percent lower than the value with $a=0$, for whatever
value may be chosen for the HS diameter of the atoms.  Thus, this virial type
effective interaction between ions and atoms can easily increase the
critical temperature, but it cannot lower it in any substantial amount.
We next studied the effects of other empirical forms for the
interactions; for example $ a \rho_{a} \rho_{i} ^\beta $ with $ \beta =
2/3$ and $1/3$, which could conceivably result from the interaction between
atoms and delocalized electrons. Again using $a$ as a variational parameter
to minimize the critical temperature, we found that the critical temperature
may be lowered by a significant amount (up to 60 percent) for $\beta=1/3$
and $a$ around -1900 a.u.. The full line of fig. 3 shows the coexistence
curve for a cesium parametrization of the metallic component and this
empirical form for the intactions with the atoms, with the optimal value of
$a$ being used. Although the shape of the coexistence curve clearly
improves compared to including the atoms only as hard spheres, the curve is
far from having the experimental shape. Moreover, the critical temperature
is still too high, even with the parameter $a$ taking the above, unphysical,
value. The optimal value for $a$ implies that, at the critical density, an
atom in the system has an energy which is 2.5 eV below that of a free
atom; in contrast, an electron ion-pair are bound to the metal at the
melting point (relative to a free atom) by less than one eV. We conclude
that such, atom-charges, effective interactions are unable to bring the
mean-field results into quantitative agreement with experimental observations.

\subsection{Criticism}

The above description of our efforts to integrate the M-nM
transition with that of the liquid-vapor one suffers from internal
inconsistencies as well. Not only did our efforts fail to yield results in
semi-quantitative agreement with observations, but also it is hard to
understand how a mean-field treatment can adequately describe the
coexistence of localized and delocalized occupation of the states
available to the valence electrons in the systems of interest. Clearly
the statistical mechanics of the treatment are a straight-forward demand for
thermal equilibrium between valence electrons localized in atoms and
coexisting delocalized valence electrons, but there is a lack of a
microscopic description which can allow for this coexistence. It seems
clear that density fluctuations are the microscopic element which permits
such a coexistence. Then, structural features can allow for valence
electron localization in atomic states or a sharing of valence electrons,
which will lead to metallic properties when the delocalization becomes
macroscopic. Occupation of the atomic valence state requires a structural
environment unfavorable to delocalization, that is a low density local
environment. In contrast, high density local environments are favorable to
valence electron delocalization on energetic grounds. On this basis, and
recalling that the data relevant to the expansion of liquid metals (a
change in coordination) is in conflict with a mean-field approach, we
decided to investigate a model which explicitly considers equilibrium density
fluctuations and their implications. The crucial point, however, is to
calculate the energy of the system not in mean field but taking into
account the strong inhomogeneities due to clustering and atom formation.
We believe that the energy of each ion can be represented as depending on
the local density of its environment; the functional dependence of this
energy is extremely important in determining the resulting material
structure and phase coexistence.

\section{model}
\label{sec:model}

\subsection{Configurational energies}

As previously noted, our main problem consists in being able to calculate
the energy of a given statistical configuration of the system.
Such a configuration will be inhomogeneous, containing regions of high
local density (clusters), in which the valence electrons are delocalized
over the region, and regions with the valence electrons localized on the
ions, $\it {i.e.}$ atoms. Clearly, on trying to calculate the energy of such
a configuration, one may not use pair interactions, as the energy in the
clusters is far from being describable in this manner. Similarly, using a
macroscopic mean-field average, with the energy as function of the
macroscopic average density, has already been discussed and found
unsatisfactory. Thus, we proceed to take into account the inhomogeneity of
the configurations in the following way.  We describe the energy of a
configuration by the sum: $ U_{c} = \Sigma u_i$, over all the ions in the
system $i=1,..N$, and look for a workable approximation for the
contribution of each ion, $u_i$.

The mean-field approach, described in the previous section, takes the
energy $u_i$ to be a function of the global density $\rho$ and is thus
equal for all the ions in the configuration, $u_i=u(\rho)$. On the other
hand, in a system with pair interactions, the exact form for $u_i$ is given
by the integral of the pair potential,
$\phi(r)$, with the local density around the ion:
$$ u_i= {1 \over 2} \int d{\bf r} \ \phi(|{\bf r}-{\bf r_i}|) \
g({\bf r},{\bf r_i}) \ \rho({\bf r}),$$
where $g({\bf r},{\bf r_{i}})$  is the radial distribution function. The
mean-field approximation for the system with pair interaction takes $
g({\bf r},{\bf r_i}) = 1$ and gives a linear dependence of the energy per ion
with the global density, $u(\rho)= \Phi_o \ \rho /2$; $\Phi_o$ is the
total integral of the interaction potential. The exact result may be
written in an intuitive way as $u_i=u_{mf}(\hat{\rho}_i)$, where
$\hat{\rho}_i$ is an effective density:
$$ \hat{\rho}_i= {1 \over \Phi_o} \int d{\bf r} \  \phi(|{\bf r}-{\bf r_i}|) \
\rho({\bf r}),$$
which describes the local enviroment of each ion. In a metal-atom fluid
the energy cannot be described as pair interactions and the
mean-field energy per ion is not a linear function of the average density.
However, we still may try to find an approximate form for the energy per
ion in terms of a local effective density.

As a preliminary step, which checks the form of this energy in a system with
mixture of localized and delocalized electronic states, we carried out
exact diagonalizations of tight-binding calculations on a finite body
centered cubic (bcc) lattice, with a single orbital per site, and
calculated the ground state energy. The samples have a partial occupation
of the lattice sites by monovalent atoms (site energy $-E$, of order the
alkali ionization potential) and the remaining sites are empty (site energy
$0$); the hopping matrix element between all nn sites ($t$) is taken to be a
constant. A spectrum of realizations were examined choosing randomly
disordered lattice-site occupations. Ensembles with a total number of 128
to 1024 sites were examined, at an occupation which ranged from nearly
empty to nearly full. Values of $E/t$ from 12 to 24 were examined. In all
cases the ensembles yield an electronic structure which depends on the
specific realization. For given $E/t$, we calculated the total ground state
electronic energy shift (from $E/t$) per occupied site for all realizations
in our ensemble. As a function of the average occupation density, this
shift has a scatter over the realizations which is substantial for low
densities, see fig.4(a). However, if the data is replotted as a function of
the average fractional occupation of nn sites to the occupied ones in each
realization, the scatter is found to be reduced by as much as a factor of
ten, see fig.4(b). The total ground state energy shift per ion for the entire
ensemble can then be described by a single curve. The data can be
equally well fitted by a single function which only depends on the fraction
of occupied sites with $n$ occupied nn ($n=0$ to $8$). Thus, the total
ground state energy of the system can be well approximated by a description
in terms of an energy $u(\hat{\rho}_i)$ per ion, with a local effective
density $\hat{\rho}_i$ which gives a measure of the coordination number of
each ion.

The independent electron tight-binding model, described above, suggests
that the energy per ion for random configurations may be described in
terms of the nn configurations of each ion. However, that model is, of
course, a very poor representation of a metal-atom fluid. For a given
environment of an ion, a local effective density, its energy should be
reasonably well described, if the valence electrons are delocalized over
the local region, by the previously discussed second order pseudopotential
calculation with the electronic exchange, correlation, and screening
effects taken into account, and assuming that the local environment is
replaced by a macroscopic one of the same density. Our proposal, then, is
to use the mean-field energy per ion $u(\rho)$, calculated as in the
previous sections, to approximate the energy per ion via
$u_i=u(\hat{\rho}_i)$; the effective density, $\hat{\rho}_i$, is to be a
simple function of the coordination number. Thus, for a given ion and
limiting the number of its nn to eight (as observed experimentally for high
densities and low temperatures), we shall take $\hat {\rho} = (n+1) \rho
_{0} /9 $ with n being the number of its nn (between 0 and 8); $ \rho _{0}$
is taken to be the density of the liquid metal at the melting point, thus
assuming that equilibrium density fluctuations are due to a change in
coordination rather than to nn distance changes.

For ions without any nn ($ n=0 $), this energy calculated as in a metal is
suspect due to the long screening length, so it is compared with that of a
free atom ($\it {i.e.}$ minus the ionization energy) and in a variational
spirit the lower value is chosen (this choice is along the lines noted in
our previous work \cite {Chacon,Hernan2}). Although the calculated energy
difference is small in this comparison, the atomic state is lower for all
the alkalis. For a single nn, the metallic energy is found to be lower than
that of the isolated dimer, for all alkalis. This approach thus includes
the structurally-based possibility of valence electron localization, to
form atoms, or some degree of delocalization. Results for u(n) calculated
in this manner, for the alkalis, are shown in fig. 5, using the Shaw
pseudopotential fitted to one bar at the melting point; an example is also
given of results arising from use of the Ashcroft pseudopotential fitted to
0.45 packing fraction at the melting point. These data have the same
functional dependence, with coordination, as the tight-binding results quoted
previously, as can be appreciated by comparing fig. 5 with the previously
noted data in fig. 4; in both cases there is a clearly nonlinear dependence
of the energy with local density, reflecting the non-additive character of
the interactions arising from valence electron delocalization.

\subsection{Structural background}

Having proposed a recipe to calculate the energy of an ion in a given
environment and thus the energy of any inhomogeneous statistical
configuration of the system, we wish to use it to self-consistently
calculate the equilibrium structures of the materials of interest at chosen
thermodynamic conditions. The simplest manner in which to implement the
calculations we propose is as follows. A theoretical construction which
will yield a material expansion through a change in the average
coordination (as observed), which simplifies the study of a fluid with
equilibrium density fluctuations, and which is known to allow for a phase
transition without broken symmetry, the liquid-vapor type, is a
model in which particles (atoms and ions) are constrained to partially
occupy the sites in a chosen lattice. Thus disorder is an intrinsic feature of
such a treatment and need not be inserted in an $ \it {adhoc}$ manner, such
as a Gaussian, or other, distribution of site energies and/or of transfer
elements, typical of Anderson model treatments. As we are interested in
treating structural, thermodynamic, and electronic properties on a similar
footing, a completely $ \it {adhoc}$ treatment of the disorder would be
inimical to our purpose.

Experimental data \cite {Winter} on the nn distance and average
coordination of coexisting liquid cesium indicates that the lattice
gas is an adequate treatment of short-range effects. The measurements show a
nn distance which changes little from the liquid at the triple point to
near-critical conditions, while expansion takes place though a changing
average coordination number. It is clear, however, that long-range effects
due to the imposition of a lattice treatment on a fluid problem will tend
to underestimate the system entropy and thus, among other effects, lead to
an overestimate of the critical temperature. Nevertheless, such an approach
seems a suitable first step. It has been noted that near-critical effects in
metal-atom fluids appear to be partially due to nonadditive effects in the
interactions \cite {Gold1,Gold2}, a feature we include in the model.
Improvements to our model will certainly be possible.

In our model, we allow the ions to partially occupy the sites of a bcc
lattice, which allows for the correct coordination of the sites when
compared to the dense, alkali, liquid metals. The lattice parameter is
determined by the condition that, at full occupation, the maximum density
be that of the liquid at the melting point, $ \rho _{0} $. The energy for
each ion is taken to be $u(\hat{\rho})$; that is, the free energy per ion
previously calculated, in the continuum mean-field description, for a
macroscopic average density $\hat{\rho}$ and setting the temperature to
zero. The equilibrium properties are then obtained using grand canonical
Monte Carlo simulations; we can then proceed to calculate the equilibrium
structures, phase coexistence, and other features of interest.

As a test of the dependence of coexistence data on the functional
dependence of $u( \hat {\rho }) $, we note the following. A lowest-order
approximation scheme, to treat the problem of interest to us, could be
based on experimental parameters only by assuming that the energy for an ion
with no nn would be the atomic energy while that for one with the full
complement of occupied neighbors (eight) would correspond to the cohesive
energy $ u_{c} $ of the liquid at the melting point. A linear interpolation
for the intermediate cases, $n$ occupied neighbors, would give the
equivalent of an Ising model, with a pair potential, at nn only, being
given by the cohesive energy of the liquid metal at the melting point
(relative to separated atoms) divided by four. The coexistence curve
and critical parameters in that case are obtainable using well-known methods
\cite{Fisher}. Using the results of the Ising model mapped to a bcc
lattice gas yields, for example, $k_{B} T_{c} = 0.79385 ( E - u_{c} ) / 2. $
The resulting critical temperatures for the alkalis would then be of order
twice the experimental ones for all the alkalis and with the observed
systematics, as can be verified by using the experimental cohesive energies
of the liquid metals at their melting point (table I). The critical
densities resulting from this approximation would then be half of
$\rho_{0}$. It should be noted that this approach gives critical
temperatures which, though high, are an improvement over the even higher
mean-field results. However, the coexistence curve would be found to be
symmetrical, in contrast to observations, and the critical densities would
be too large. The decrease in the critical temperature of this Ising model,
when compared to macroscopic mean-field results, is due to clustering
induced by the pair interactions. The symmetry of the coexistence curve
arises from the symmetry between occupied and empty sites. Such an
approximation scheme fails to take into account many-body effects due to
valence electron delocalization, which yield the concave-up u(n) vs n
curves. The enhanced cohesion of intermediate clusters, implied by such
curvature, further enhances clustering. However, the curvature implies that
clusters may be more strongly or more weakly bound, relative to nearby
local densities, than the above Ising approach. The local tangent to the
u(n) curve is an effective Ising coupling at that local density; for low n
this coupling is stronger than that due to connecting the end points, and
for large n it is weaker. Thus, although the curvature yields more
clustering, the critical temperature due to a model with such curvature
need not be higher than that of the straight-line Ising model. Also, the
curvature breaks the symmetry between occupied and empty sites. Both
effects could yield improvements in calculated data, allowing closer
approach to observations. We have found that this is indeed the case. We
proceeded to discuss the test of our proposed scheme on various alkalis and
a spectrum of their properties.

\subsection{Monte Carlo Simulations}

We will present the results of grand canonical Monte Carlo simulations
for this model, applied to the alkalis from cesium through sodium. First,
with an energy per ion taken to be $u( \rho)$, a mean-field lattice
treatment, we have compared the critical data with that of the previous,
continuous, metallic model to explore the influence of the lattice. The
results are given in table II (labeled MF-lg) and are found to be quite
close to those of the continuous model; this is due to the low critical
density resulting from the mean-field approximation, which makes the
entropy of the reference system close to the ideal-gas entropy. However,
the crucial point is to take the inhomogeneities in the statistical
configurations into account. Simulations were then performed using the
energy per ion $u(\hat{\rho})$; this function has been previously discussed
and is shown in fig.5. We have used a simulation cube with twelve bcc cells
on each side (3456 sites) and periodic boundary conditions. Some results
have been checked using a larger cube (8192 sites), without appreciable
changes in the results. The simulations, carried out at fixed temperature
and chemical potential, give the equilibrium density and internal energy of
the system. The pressure is obtained by thermodynamic integration. All
results which follow are obtained using the Shaw pseudopotential fitted to
1 bar of pressure at the melting point.

\subsubsection{Coexistence}

Calculations of the coexistence curves of the alkalis were carried out.
Once again we first examine the critical parameters as an approximate
test of the results, before continuing to a more detailed comparison with
observations. The critical parameters obtained are also given in table II and
labeled as MC. The critical temperatures from our model are higher (by
about 400 K) than those of the real fluids, a result one might expect from
the lattice-gas treatment of the configurational entropy. However, the results
in table II show that the correlation effects included in the present model
produce very large decreases of the critical temperatures compared with the
previously mentioned mean-field approach, and also compared to the Ising
approximation. Also, the critical densities and pressures are brought into
reasonable agreement with experiments. Moreover, the relative critical
parameters of the various alkalis, in our model, are very similar to the
relative experimental ones. These results, it should be noted, also exhibit
quantitative improvements relative to those, using the Ashcroft
pseudopotential, reported in the table of our letter \cite {Tarazona}.
Finally, they are used to scale the coexistence curves reported below.

In fig. 6, we show the coexistence curves, in critical reduced units,
calculated and observed for cesium, this data should be compared with that
shown in fig. 1 of our previous work \cite {Tarazona}. Using the present
methods, the shape of the coexistence curve is in good agreement with
experiments. It recovers the strong liquid-vapor asymmetry. The figure also
presents the diameter function $ \rho _{d} = ( \rho _{L} + \rho _{V} ) / 2
\rho _{C} $. The errors in the simulations, near critical point, prevent us
from a detailed analysis of the deviation of this function from a linear
law, observed in the experimental results \cite {Jungst}.  We have also
calculated the coexistence curves for other alkalis. The results are shown
in fig. 7; in reduced critical units, they are all very similar. In
agreement with experiments, our results for different alkali fluids give
similar reduced phase diagrams. The accuracy in our critical density
determinations is poor and does not allow for a discrimination of
the small differences observed \cite {Hohl,Jungst} in the shape of the
coexistence curves. In our calculations, we estimate an uncertainty of $
\pm 10$ percent in $ \rho_{c} $ and half that value for $ T_{c} $; the
relative values among the alkalis should be reliable, as they are all
estimated in the same manner. These relative values show the experimentally
observed systematics, as can be seen from the comparison shown in table II.

There are no qualitative changes between the coexistence curves reported
here and those, using the alternate Ashcroft pseudopotential, of our
previous work  \cite {Tarazona}. However, quantitatively, for example, the
critical temperatures are lowered, approaching observations, on using
the present rather than the previous approach. The reduction of the
critical temperatures on using the Shaw, rather than Ashcroft,
pseudopotential is about 1200 K, leaving our results about 400 K
too high when compared to experiment. We believe the relative effect is due
to the reduction of the discrepancy between the calculated and observed
cohesive energies of the liquid metal at the melting point (see table I), the
general shape of the u(n) curves staying generally invariant, as can be
seen by the example shown in fig. 5.

\subsubsection{Structural features}

Gross structural aspects such as a nearly-linear decrease in the average
coordination of ions in the expanded, coexisting, liquid result from this
calculation, see fig. 8. The mean-field result (dotted line) underestimates
the mean coordination number obtained on approaching the critical
point, this is an effect of clustering. The analyses of experimental neutron
scattering data \cite {Winter}, along the liquid line at coexistence,
yield pair correlation functions which show a mean-field linear behavior on
reducing the density from the triple point to a density of about one-half
of that value. Such data, at even lower densities, are difficult to analyze,
due to the broadness and lack of definition of the first peak in the
deduced pair correlation function, and, so far, there is no strong evidence
for clustering from that data. On the other hand, inverse MC calculations
\cite {Nield}, based on the experimental structure functions themselves, do
show evidence of clustering, even though such calculations tend to
underestimate clustering effects because of the averaging inherent in the
method. Also, dimerization effects have been invoked to analyze magnetic
susceptibility measurements \cite {Freyland}, though such an analysis is
more speculative than the results of the inverse MC. If low density
structural measurements are possible, we await their results for further
comparison with our theoretical predictions.

Other structural features which result from our calculations are
exemplified by data such as that given in fig. 9. The figure shows the
fraction of occupied lattice sites within clusters of a given size, as a
function of that size. A cluster here is defined as a set of occupied sites
which have no nn outside the cluster and which are connected by nn within the
cluster. The figure is calculated for low density cesium vapor, at the
critical temperature obtained from the model and at two average density
values: 0.023 and 0.10 $\rho _{o}$. The smaller density is well below the
critical density and the obtained probability of finding a large cluster
decays exponentially with the cluster size. The larger density is near the
percolation threshold and the calculated probability of a given cluster
decays only as a power of the cluster size. Effects due to such percolation
will be discussed later in connection with electrical conductivity
estimates. The fraction of the total number of occupied sites in clusters
of a given size decreases with the cluster size at fixed average system
density. The scatter of points results from data due to various
realizations of the MC simulations: small clusters have high
probability, they are obtained in most of the realizations and thus
show little scatter; large clusters are improbable, they are only
observed in some realizations, and thus show a substantial scatter.
This latter effect is due to the limited size of the simulation box. Note the
reduction in the number of atoms, clusters of size one, as the average
system density is increased; there is no jump in this number with
density variations, for temperatures above the phase coexistence.

\subsubsection{Electrical Conductivity}

We have next explored electronic transport properties related to structure.
The experimental signature of a M-nM transition in these systems is a
decrease of several orders of magnitude in the conductivity of the expanded
fluid. In our model, such behavior is driven by the percolation of the
ionic cluster structure, which can be interpreted as leading to macroscopic
delocalization of valence electrons, rather than by considerations such as
those of Mott or Anderson. Such percolation, however, need not be unrelated
to the physical basis in the Mott and Anderson pictures. From typical
configurations of our MC simulations, we have obtained the cluster
structure at different temperatures and pressures. The electrical
conductivity was then estimated following the Kirchoff's law model proposed
by Nield et al \cite {Nield}, with a fitting of the experimental
conductivity at the maximum density.

The details of the approach we have used are the following. For a given
equilibrium configuration of the occupied sites in our simulation box,
nn occupied sites are replaced by a bond resistance of fixed value, no other
effects are taken into account. An electrical potential difference is then
assumed to be applied to two parallel faces of the box, while periodic
boundary conditions are applied to the other sets of box faces. The
potential difference would cause a current to flow, through the resistor
network, using Ohm's law and requiring that the net current into each site
must be zero (Kirchoff's law). The equations are then solved for the
current as a function of the potential difference and, taking into account
the box size, the conductivity is calculated as a function of the assumed
bond resistance. Appropriate averages are carried out for a set of
realizations of the simulation at each set of conditions. Clearly a finite
conductivity is only found if there is at least one bond-percolation
cluster traversing the potential difference. These conductivity estimates
are intended to probe the effects of the obtained cluster structure on the
transport properties. The estimates are based on an imposed macroscopic
treatment of each nn bond in the material, a resistor, instead of a direct
quantum treatment. At first glance such an approach appears to be a gross
simplification, yet there is some reassurance: the bond resistance required
to fit the experimental data at high densities is 15000 $ \Omega $, a
result rather similar to the quantum of resistance, without dissipation,
per electron, in a single channel ($ h/(2 e^2) $ ): 12906 $ \Omega $; the
appropriate classical resistance should indeed be greater than the quantum of
resistance, due to a transmission coefficient effect. Further justification
of this approach is our conjecture that the electron mean free path is
rather short in the physical cases, at the conditions considered, so that
direct transport between sites separated by distances larger that that of
nn sites would only yield negligible contributions.

The results in fig. 10 show the conductivity estimated in this manner along
the obtained critical isotherm, for Cs, versus the pressure. They are in
excellent agreement with the experimental observations \cite {Gener1}, as
is shown in the figure. The figure also shows the calculated conductivity
versus density at the critical temperature, and that which would arise from a
random occupation of the lattice. The difference clearly shows the effects
due to energy-driven clustering. The percolation density (at which the
conductivity goes to zero) at $ T _{c} $ is less than half of that obtained
with random occupation. These results are almost unchanged from those
reported in fig. 2 of our letter \cite {Tarazona}, which arose from the
Ashcroft pseudopotential. A line delimiting the densities and temperatures
for the percolation onset is shown in fig. 6, accompanying our coexistence
calculations for Cs. It is clear that this line does not intersect the
coexistence curve at the critical point, as has been speculated. The
results for the percolation line are similar in the spectrum of alkali
fluids.

\subsubsection{Summary}

We have presented a lattice-gas implementation of a model allowing a unified
study of the structural, thermodynamic, and electronic properties of
metal-atom fluids. The model takes into account the inhomogeneity
of statistical configurations of the system. Nonadditive interactions,
due to valence electron delocalization, are included. A self-consistent
procedure is used to determine the equilibrium structures: a MC simulation
which goes beyond mean-field, as is required. Although the model is a very
simplified representation of a metal-atom fluid, comparisons of results
obtained, for the alkali family and a spectrum of data, show that the model
contains the basic ingredients to allow reproduction of the peculiar
behavior observed in these systems. These peculiarities include the
M-nM and liquid-vapor transitions and the connection between ionic and
electronic structures. Our results for the scaled coexistence curves are in
good agreement with  observations. The calculated critical temperatures are
still somewhat high and the critical pressures low, the densities being
adequate. It is reasonable to expect that, with a more realistic
(non-lattice) description of the fluid entropy, a similar model (though
more cumbersome to study) would give a quantitatively good result for all
system properties. Our simplified model appears to allow a unified
understanding of the peculiar characteristics of the alkali fluids. The
model also shows the similarities and differences between these materials
and pairwise interacting ones.

\section{Conclusion}

This paper reports investigations of theoretical models which attempt to
probe the liquid-vapor and M-nM transitions in the alkali fluids from a
unified point of view. Mean-field attempts were investigated first, as an
extension of previous work. Although, based on statistical mechanics, they
are capable of unifying the two transitions, they are lacking a microscopic
basis for treating the coexistence of occupation of the valence electron
states localized in atoms and those shared by various ions, which
eventually result in accounting for metallic properties. Further, such
models do not yield semi-quantitative agreement with observations of the
structure and coexistence curves of alkali fluids.

Our model was then discussed. It includes treatment of the equilibrium density
fluctuations driven by interactions, which include nonadditive effects due
to, at least, partial valence electron delocalization. This model allows a
unified, self-consistent, treatment of structural, thermodynamic, and
electronic properties. The lattice-gas implementation of this model via a MC
simulation method was then undertaken. The results of such a treatment do
reproduce the spectrum of peculiar features observed in alkali fluids,
including the coupling of structural and electronic properties. It
should be emphasized that except for the nonadditive effects, due to
valence electron delocalization, the model is no different from that
applicable to insulating fluids. Thus we no longer consider that the
treatment for both types of materials is intrinsically different, except
for effects due to electron delocalization. We believe that our model
contains the basic ideas required for an adequate treatment of such fluids.

The energy recipe we have used is a strong simplification of nature,
treating electrons which are delocalized over various ions as
macroscopically delocalized and also neglecting thermal effects. Further,
although our calculations with the model treat the short-range structure of
the material in a manner we believe to be reasonably adequate, the
long-range structure is oversimplified by the imposition of a lattice.
Improvements are required for a full quantitative treatment. The main
features of such improvements should seek to give a better treatment of
entropy effects with no lattice-like effects in the long-range fluid
structure. Such an improvement should be treatable in an improved,
non-lattice,  MC simulation for a fluid. We urge that such a calculation be
undertaken. Also, it would be interesting to attempt an extension of the
ideas we have presented in order to seek an improved understanding of
polyvalent atoms; mercury can clearly be the testing ground of such
extensions. Finally, attempts to integrate Anderson and Mott treatments
with the work presented here could yield a deeper understanding of
metal-atom fluids. We hope our calculations are a useful stepping-stone
to allow guidance of experimental efforts and that extensions of this
work will allow theoretical extrapolations to materials at experimental
conditions other than those achievable with static techniques.

\acknowledgements

We thank Jianping Lu for some computations related to the tight-binding
calculations. Our work was partially supported by the Direcci\'on General
de Investigaci\'on Cient\'{\i}fica y T\'ecnica (Spain) under Grant
PB91-0090 and the NATO Office of Scientific Research via grant
SA.5-2-05(CRG.940240). One of us (JPH) is also grateful for financial
support to the W. R. Kenan, Jr. Foundation, the Spanish Ministry of Science
and Education, and the Instituto Nicolas Cabrera.

\newpage
\begin{figure}
\caption{ Difference between the mean-field free energy of Na (Ashcroft
pseudopotential) with a HS reference system and that with the OCP (eV) vs
density (a.u.), for various temperatures: solid line, $ T=2000 K $;
long-dashed line, $ T=4000 K $; short-dashed line, $ T=6000 K $.}
\label{fig1}
\end{figure}

\begin{figure}
\caption{Coexistence curve obtained for metallic rubidium in the
mean-field approximation, with a HS reference system. Solid line: Shaw
pseudopotential, dotted line: Ashcroft pseudopotential. Temperature in K,
density in a.u..}
\label{fig2}
\end{figure}

\begin{figure}
\caption{Coexistence curves obtained for metallic cesium in equilibrium with
its atoms, mean-field treatment: The dashed line shows the results on
treating the atoms as hard spheres, with a diameter ($ 10 a.u. $) different
than that of the ions ($ 8.80 a.u.$), but not otherwise interacting with
the charges; for this case the temperature scale should be doubled. The
full line present the results when, in addition, an empirical interaction
with the charges ($ a \rho _a \rho_i ^{ \beta} $) is included. The
parameters, $ a = -1900 a.u.$ and $ \beta = 1/3 $, are tuned to minimize
the critical temperature. The open points show the experimental coexistence
curve of  cesium (ref.[3]). }
\label{fig3}
\end{figure}

\begin{figure}
\caption{  Ground state, tight-binding, energy shifts per ion ($\Delta
E/t$) calculated for a set of random occupation realizations in a bcc
lattice, with $E/t =24$: (a) as function of the average occupation density,
(b) as a function of the average occupation of a nn site for each occupied
site in the various realizations.  }
\label{fig4}
\end{figure}

\begin{figure}
\caption{ Energy per ion in the alkalis, with the Shaw pseudopotential,
$u(\rho)$ vs density normalized to density of the liquid metal at
the melting point $\rho _{0}$. Solid line: cesium ($ \rho _{0}=1.84 g
/ cm^3 $), dotted line: rubidium ($ \rho _{0}=1.47 g / cm^3 $),
dashed line: potassium ($ \rho _{0}=0.828 g / cm^3 $), and
dashed-dotted line: sodium ($ \rho _{0}=0.924 g / cm^3 $). The
crosses at the end each curve are the experimental cohesive energies.
For comparison, the circles show the results for Cs using the Ashcroft
pseudopotential, which were used in ref. [5]. Energies are in eV. The
$\rho$/$\rho_{0}$=$1/9$ (i.e., n=0) values are experimental data for
the free atoms. }
\label{fig5}
\end{figure}

\begin{figure}
\caption{ Liquid-vapor coexistence curve of cesium, in reduced critical
units. Filled circles: present MC simulation; full line: fit to the
experimental results of ref.[3]. The MC simulation diameter function $
\rho_{d} $ is also plotted (crosses). The triangles, joined by a line to
guide the eye, delimit the region of cluster percolation. Note the slight
differences with fig.1 of ref. [5], due to a change in the pseudopotential
choice.}
\label{fig6}
\end{figure}

\begin{figure}
\caption{Calculated liquid-vapor coexistence curves for the alkali fluids,
using the Shaw parametrization, in reduced critical units; circles: cesium,
squares: rubidium, triangles: potassium, and stars: sodium.}
\label{fig7}
\end{figure}

\begin{figure}
\caption{Average coordination number versus normalized density, along the
the liquid branch of the coexistence curve of cesium, resulting from the MC
treatment of the lattice gas. The dotted line shows the mean-field result,
with average coordination number equal to $8 \ \rho/\rho_0$. }
\label{fig8}
\end{figure}

\begin{figure}
\caption{Fraction of ions in clusters of each cluster size (number of ions)
in a log-log plot. The calculations use the MC lattice-gas treatment, for
cesium at the critical temperature, at two average densities: $ 0.023$
$ \rho _{o}$ (solid points) and $0.10$ $ \rho _{o}$ (open points). }
\label{fig9}
\end{figure}

\begin{figure}
\caption{ On the left we show the electrical conductivity $ \sigma $ for
cesium versus the reduced pressure along the critical isotherm. Crosses
from our model, open circles are experimental values from ref [1], both
normalized to the conductivity $ \sigma _{0} $ at our highest density $
\rho _{0} $. On the right we show our calculated electrical conductivity
but now versus the normalized density; points are our results for $ T = T
_{C}$, and triangles for the random occupation of the lattice. Note the slight
differences with fig. 2 of ref. [5], due to a change in the
pseudopotential choice.}
\label{fig10}
\end{figure}

\newpage
\widetext
\begin{table*}[h]
\noindent
\label{tableI}
\caption{Comparison of the calculated results for the liquid alkali metals,
at their melting point, using the Ashcroft and Shaw pseudopotentials (the
core radii marked by an asterisk are fitted to a 0.45 packing fraction for
the liquid metal at the melting point). Core radius ($r_{c}$) and
hard-sphere diameter ($d_{HS}$) in a.u., packing fraction ($\eta$) and
excess entropy ( $ S_E/k_B $ )  are dimensionless, atomic energy ($E$)
and liquid cohesive energy ($u$) (referred to ion cores and valence
electrons at infinite separation) in eV, conductivity ($\sigma$) in $ 10^5 (
\Omega  cm)^{-1} $,  and pressure ($P$) in bar. Note the systematic
improvement in $u$, and to a lesser extent in $E$, on fitting to $P$ --
Shaw rather than to  $\sigma$ (or packing fraction) -- Ashcroft.}

\begin{tabular}{|c|cccccccc|cccccccc|} \hline & \multicolumn{8}{c|}
{ CESIUM } & \multicolumn{8}{c|}{ RUBIDIUM } \\
& $ r_{c} $ & $d_{HS}$ & $\eta$ &  $E$ & $u$ & $ \sigma $ & $ S_E/k_B
$ & $ P $ &
$ r_{c} $ & $d_{HS}$ & $\eta$ & $E$ & $u$ & $ \sigma $ & $ S_E/k_B $
& $ P $ \\
\hline
 & & & & & & & &  \\
 Ashcroft  & $2.62^*$ & 8.80 & 0.45 & -3.89 & -5.12 & 0.70 & 3.81 & -6260
           & $2.40^*$ & 8.19 & 0.45 & -4.23 & -5.39& 0.53 & 3.81 & -7380 \\
 Shaw      & 4.73 & 8.82 & 0.44 & -3.93 & -4.73 & 0.68 & 3.79 & 1.0
           & 4.30 & 8.18 & 0.44 & -4.16 & -5.05 & 0.64 & 3.77 & 1.0   \\
 Exp       & -- & -- & -- & -3.89 & -4.69 & 0.28 & 3.56 & 1.0
            & -- & -- & -- & -4.18 & -5.01 & 0.45 & 3.63 & 1.0   \\
 & & & & & & & &  \\
\hline
\end{tabular}

\begin{tabular}{|c|cccccccc|cccccccc|} \hline & \multicolumn{8}{c|}
{ POTASSIUM } & \multicolumn{8}{c|}{ SODIUM } \\
& $ r_{c} $ & $d_{HS}$ & $\eta$ &  $E$ & $u$ & $ \sigma $ & $ S_E/k_B
$ & $ P $ &
$ r_{c} $ & $d_{HS}$ & $\eta$ & $E$ & $u$ & $ \sigma $ & $ S_E/k_B $
& $ P $ \\
\hline
 & & & & & & & &  \\
 Ashcroft  & 2.12 & 7.51 & 0.42 & -4.48 & -5.76 & 0.60 & 3.40 & -9230
           & 1.70 & 6.17 & 0.44 & -4.86 & -6.61 & 1.0 & 3.80 & -10700  \\
 Shaw      & 3.92 & 7.60 & 0.43 & -4.37 & -5.36 & 0.63 & 3.65 & 1.0
           & 2.92 & 6.11 & 0.43 & -4.97 & -6.45 & 0.63 & 3.50 & 1.0    \\
 Exp       & -- & -- & -- & -4.34 & -5.26 & 0.77 & 3.45 & 1.0
            & -- & -- & -- & -5.14 & -6.25 & 1.0 & 3.45 & 1.0    \\
 & & & & & & & &  \\
\end{tabular}
\end{table*}

\narrowtext

\begin{table}[h]
\noindent
\label{tableII}
\caption{Estimated critical conditions: using continuum mean-field with
hard-sphere (MF-hs) and lattice-gas (MF-lg) reference system entropy, Monte
Carlo (MC) simulation for the present theory (Shaw pseudopotential), and
the experimental results of ref. [2] and [3]. The temperature $ T $ is in
Kelvin, the pressure $ P $ in bar, and the density $ \rho $ in $ g $ $
cm^{-3} $. The MC results are later used to normalize the results presented
in various figures.}

\begin{tabular}{|c|ccc|ccc|} \hline & \multicolumn{3}{c|}{ CESIUM } &
\multicolumn{3}{c|}{ RUBIDIUM } \\
& $ T_{c} $ & $ \rho_{c}$ & $ P_{c} $ & $ T_{c} $ & $ \rho_{c}$ & $ P_{c} $ \\
\hline
 MF-hs  & 6330 & 0.022 & 6.7 & 6820 & 0.019 & 9.3    \\
 MF-lg  & 6540 & 0.030 & 8.3 & 7060 & 0.025 & 11.9    \\
 MC & 2350 & 0.47 &  60 & 2475 & 0.35 & 73    \\
 Exp & 1924 & 0.38 & 92.5 & 2017 &0.29& 124.5   \\
\hline
\end{tabular}

\begin{tabular}{|c|ccc|ccc|} \hline & \multicolumn{3}{c|}{ POTASSIUM } &
\multicolumn{3}{c|}{ SODIUM } \\
& $ T_{c} $ & $ \rho_{c}$ & $ P_{c} $ & $ T_{c} $ & $ \rho_{c}$ & $ P_{c}
$ \\
\hline
 MF-hs  & 7310 & 0.011 & 13.0 & 9020 & 0.014 & 34.6    \\
 MF-lg  & 7580 & 0.014 & 16.2 & 9390 & 0.018 & 45.4    \\
 MC  & 2550 & 0.22 &  70 & 2970 & 0.22 & 128    \\
 Exp & 2178 & 0.18 & 148 & 2485 & 0.30 & 248   \\
\end{tabular}
\end{table}

\end{document}